# Is there currently a scientific revolution in scientometrics?

Lutz Bornmann

Division for Science and Innovation Studies

Administrative Headquarters of the Max Planck Society

Hofgartenstr. 8,

80539 Munich, Germany.

E-mail: bornmann@gv.mpg.de

**Abstract**

The author of this letter to the editor would like to set forth the argument that scientometrics is currently in a phase in which a taxonomic change, and hence a revolution, is taking place. One of the key terms in scientometrics is scientific impact which nowadays is understood to mean not only the impact on science but the impact on every area of society.

For some years now there has been a widely accepted taxonomy in scientometrics in order to use publication and citation data to investigate certain issues. Taxonomy is here understood to be a roughly outlined scheme used throughout the community of scientometricians for the purposes of research and its application (Wray, 2011). We can describe the recent decades in scientometrics as Kuhn (1962)'s normal science in a mature field of research which has seen refinements, reworkings and enhancements, but no fundamental changes to the taxonomy. For example, the indicators used as a rule in scientometrics for the evaluation of research are normalised for both the subject category and the year in which a publication has appeared. Co-citation analysis is usually used to identify research fronts in disciplines. The analysis of co-authorships is used to investigate collaboration between research groups, universities and countries.

According to Kuhn (1962) scientific revolutions can be characterised as taxonomic changes in a mature research field (Wray, 2011). The Copernican revolution is an example of a taxonomic change (further examples of revolutionary changes can be found in Wray, 2011). Whereas Ptolemaic astronomers did not consider the earth to be a planet, Copernican astronomers did. When according to Kuhn (1962) competing taxonomies do not categorise phenomena in the same way and the meanings of key terms are incompatible, a revolution in the field replaces one taxonomy with another (Wray, 2011). The author of this letter to the editor would like to set forth the argument that scientometrics is currently in a phase in which taxonomic change and therefore a revolution is taking place. One of the key terms in scientometrics is scientific impact, which is usually measured using citations in literature databases (such as the Web of Science, Thomson Reuters). Citations are seen as a proxy for quality able to measure one aspect of scientific quality – impact – (the other aspects are accuracy and importance of research) (Martin & Irvine, 1983).

Nowadays, however, impact is increasingly understood in a broader way which implies not only scientific but also many different kinds of impact. The meaning of the key term "impact" has changed primarily because information about direct impact of science in other areas of society is expected. For example, the UK funding bodies have decided that the overall framework for assessing impact in the 2014 Research Excellence Framework (REF) is as follows: "The impact element will include all kinds of social, economic and cultural benefits and impacts beyond academia, arising from excellent research, that have occurred during the period 1 January 2008 to 31 July 2013" (Higher Education Funding Council for England, 2011, p. 4). Kuhn (1962) calls these changes in the meaning of key terms 'meaning-incommensurabilities' between two different taxonomies (Wray, 2011). Today, one understands the impact of research to mean societal impact, which is not limited to science but also encompasses social, cultural, environmental and economic impact (Bornmann, 2012). In scientometrics the change of meaning will lead to changes in the practice of impact measurement (topic-incommensurability) (Wray, 2011). A scientometrician today has to deal increasingly with the different kinds of impact which need to be measured; in the past it would have been clear to that scientometrician that scientific impact should be measured with the accepted methods and instruments. New tools for measuring impact will play an important role in measuring societal impact. These tools are frequently grouped under the term 'altmetrics'. Tools, such as ImpactStory or Altmetric, measure the impact of scientific papers beyond science.

There have of course been other developments in scientometrics over recent years, which have excited great interest. The development of the h index can be interpreted as a scientometric development of this nature. The introduction of this indicator led to an enormous amount of research (a new research front using the h index was established). Unlike the change in the key term 'impact', the h index was not associated with a basic taxonomic change in scientometrics. According to Wray (2011) changes in techniques and practices can be significant contributions to a field; but we need not treat all these changes as revolutionary.

Today, scientometricians are faced with the questions of what exactly is meant by societal impact and how can it be measured. These questions can be answered in a coming period of normal science.